# Cooperative Access Networks: Optimum Fronthaul Quantization in Distributed Massive MIMO and Cloud RAN

*(Invited Paper)*

Alister Burr, Manijeh Bashar and Dick Maryopi
Dept of Electronic Engineering, University of York
Heslington, York YO10 5DD, U.K.
alister.burr@york.ac.uk

*Abstract*—We consider cooperative radio access network architectures, especially distributed massive MIMO and Cloud RAN, considering their similarities and differences. We address in particular the major challenge posed to both by the implementation of a high capacity fronthaul network to link the distributed access points to the central processing unit, and consider the effect on uplink performance of quantization of received signals in order to limit fronthaul load. We use the Bussgang decomposition along with a new approach to MMSE estimation of both channel and data to provide the basis of our analysis.

I. INTRODUCTION

In recent years, and especially in the context of the development of 5G standards, a number of novel proposals have been made for the architecture of next generation radio access networks (RANs), in order to achieve the enormous increase in user capacity-density that are predicted for 5G and beyond. Two apparently contrasting proposals that have attracted much attention are massive multiple input, multiple output (MaMIMO) [1] and Cloud RAN (C-RAN) [2]. MaMIMO proposes to serve increased numbers of users by concentrating both antennas and signal processing at a relatively small number of (non-cooperative) base stations provided with extremely large arrays, while C-RAN would distribute the antennas across the service area, but concentrate signal processing facilities in cloud-based processing centers (hence the name). More recently, however, the concept of distributed or "cell-free" MaMIMO [3] (here D-MaMIMO) has been introduced, in which the antennas are distributed across the service area. This effectively abolishes the concept of the cell (resulting in the term "cell-free") because users are now served by multiple antenna sites. Note that centralized MaMIMO retains some of the disadvantages of conventional cellular architectures, in that there are still "cell-edge" users which have poorer channels than users near the base station, and which may be more subject to intercell interference due to *pilot contamination* resulting from pilot sequences shared with adjacent cells [1].

The resemblance of the D-MaMIMO and C-RAN concepts is obvious, as is their similarity to the earlier "Network MIMO" or coordinated multipoint [4]. In fact one of the main benefits of both schemes is that they enable a highly scalable implementation of cooperation across multiple access points, as discussed in [5]. The differences between them relate mainly to the different emphases which arise from their origins: C-RAN originates in the desire of network operators to increase the efficiency and cost-effectiveness of their networks, while D-MaMIMO has a more academic origin, generalizing the approach of centralized MaMIMO. It has the advantage for our current purpose of providing a comprehensive mathematical model of the network suitable for fundamental analysis from a physical layer point of view, which we will use in the remainder of the paper. It also allows a fundamental comparison (as in [6]) of D-MaMIMO and a further candidate for the next generation network architecture: small cells.

However both C-RAN and D-MaMIMO encounter the major challenge of connecting the antenna sites (which we will refer to as *access points*, AP) to the central processing unit (CPU – also referred to in C-RAN as the *baseband unit*, BBU). In this paper, following the C-RAN terminology, we will refer to these connections as constituting the *fronthaul* network.

The implementation of this is a particular challenge on the uplink network, since it must carry the signals received at the APs to the CPU. When transferred in digital form this requires a capacity for the fronthaul links many times the corresponding user data rate. In the C-RAN literature this has been estimated as 20-50 times the corresponding data rate [7], implemented using the CPRI standard [8], typically over optical fiber. So far the vast majority of work on D-MaMIMO has not addressed the fronthaul load issue in detail, for the most part assuming that it is unlimited for the purpose of the performance analysis. It is reasonable to assume, however, that the fronthaul network will carry quantized signals, at least in the uplink direction, and that this will affect the network performance.

This paper therefore provides an approach for the analysis of the effect of fronthaul quantization on the uplink of D-MaMIMO. While there has been significant work in the context of network MIMO on compression techniques such as Wyner-Ziv coding for interconnection of base stations, here for simplicity (and hence improved scalability) we assume simple uniform quantization. We make use of the Bussgang

The work described in this paper was supported in part by UK EPSRC under grant EP/K040006, and in part by the European Commission under the MSCA-ITN programme, contract no. 675806 "5G-AURA".

decomposition [9] to model the effect of quantization. We exploit the schemes in [10], [11] to derive the optimal step size of the quantizer. In [12], the present authors propose to use a quantizer with a fixed step size to model the effect of the quantization. However, in [13], we extend our works in [12], [14]-[16] to a limited-fronthaul cell-free massive MIMO system using the Bussgang decomposition.

The remainder of the paper is structured as follows. In Section II we describe the network model, based, as discussed above, on the D-MaMIMO model of [3]. Section III is devoted to the model of the uniform quantization process, using the Bussgang decomposition, along with minimum mean square error (MMSE) estimation of both channels and signals at the CPU. This is used in Section IV to analyse both channel estimation and data detection using MMSE, and hence to obtain numerical results on the uplink performance of D-MaMIMO. Finally we discuss the implications of our results for more general cooperative radio access network architectures, including C-RAN.

## II. NETWORK MODEL

Fig. 1 illustrates the network model we will use for the D-MaMIMO system. $K$ user terminals (UTs) are served by $M$ APs over wireless links, which in turn are connected to a CPUs via the fronthaul network. The links on this network are assumed to be digital, and (for the purposes of the current paper) error-free. The received uplink signals at the APs are converted to complex baseband form, and their in-phase and quadrature components are filtered using a matched filter and sampled at a rate of one sample per symbol, and quantized with a precision of $l_{fr} = \log_2(L)$ bits each. In this paper we do not define the medium over which the fronthaul network operates: optical fiber, wireless and copper wire (using digital subscriber line – DSL – technology) have been proposed. Here we assume that APs are equipped with a single antenna each. They are distributed randomly and uniformly over a square service area of side $l_{serv} = 1$ km.

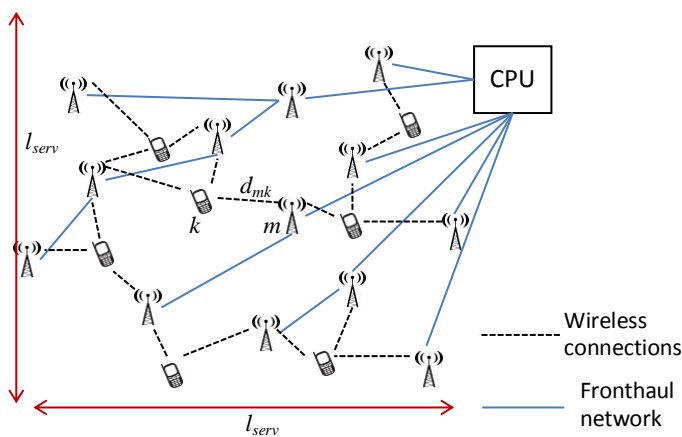

Fig. 1. D-MaMIMO network model

The wireless channel between the $k^{th}$ UT and the $m^{th}$ AP is assumed to be affected by a path loss related to the distance $d_{mk}$ between them, a log-normally distributed shadow fading term, and a frequency-flat Rayleigh fading term. Both the Rayleigh and the shadow fading terms are uncorrelated between different UTs and APs. Note that in practice a much broader band signal would be likely to be used, based on multicarrier modulation: then the fading for each subcarrier could be modelled as flat fading. As pointed out in [1] frequency domain correlation would have advantages for channel estimation in such a multicarrier system: however we do not consider this effect here.

Thus the channel propagation coefficient between the $k^{th}$ UT and $m^{th}$ AP is given by:

$$g_{mk} = h_{mk}\sqrt{\beta_{mk}} = h_{mk}\sqrt{10^{\xi_{mk}/10} \text{PL}(d_{mk})} \quad (1)$$

where $h_{mk} \sim \mathcal{CN}(0,1)$ accounts for Rayleigh fading and $\xi_{mk} \sim N(0,\sigma_{sh})$ for shadow fading with standard deviation $\sigma_{sh} = 8$ dB in this paper. Path loss is modelled using a three-slope model, as in [3]:

$$\text{PL}(d) = \begin{cases} 1 & d < d_0 \\ \left(\dfrac{d}{d_0}\right)^{-\gamma_0} & d_0 \le d < d_1 \\ \left(\dfrac{d_1}{d_0}\right)^{-\gamma_0}\left(\dfrac{d}{d_1}\right)^{-\gamma_1} & d \ge d_1 \end{cases} \quad (2)$$

with $d_0 = 10$ m, $d_1 = 100$ m, $\gamma_0 = 2$ and $\gamma_1 = 3.5$.

In this paper we assume as in [3] that the shadow fading and path loss components (jointly modelled as $\beta_{mk}$) are sufficiently stable to be accurately estimated at the APs, and that their values need be transmitted to the CPU across the fronthaul network sufficiently infrequently that the additional fronthaul load is negligible.

We also define the receive noise power at an AP in terms of the signal to noise ratio observed over a link of length $l_{serv}/2$ (for example at an AP at the center of the service area from a UT at the edge). Then the noise variance:

$$\sigma_n^2 = \left(\dfrac{d_1}{d_0}\right)^{-\gamma_0}\left(\dfrac{l_{serv}}{2d_1}\right)^{-\gamma_1}\dfrac{1}{\text{SNR}_{edge}} \quad (3)$$

Data is transmitted over the wireless channel in packets whose length is assumed to be sufficiently smaller than the channel coherence time such that Rayleigh fading can be treated as constant. For the purposes of channel estimation (of $\beta_{mk}$ over the long term and $h_{mk}$ in each packet) a pilot sequence of length $\tau$ symbols are transmitted in each packet. Here we assume that the pilot sequences assigned to all UTs are orthogonal. (Note that in practice it is only necessary to ensure that the sequences for all the UTs received at each access point are effectively orthogonal, and algorithms exist to perform the required assignment [3]). Here we assume for simplicity that the received pilot sequences are transmitted over the fronthaul and that $h_{mk}, \forall m,k$ is estimated at the CPU.

## III. FRONTHAUL QUANTIZATION

### A. Quantizer model

We assume that the in-phase and quadrature components of the received uplink signal at each AP are uniformly quantized using an $L$-step (with $L$ even) uniform quantizer with step size $\Delta$ defined as:

$$g(x) = \begin{cases} -\frac{L-1}{2}\Delta & x \leq -\left(\frac{L}{2}+1\right)\Delta \\ \left(l+\frac{1}{2}\right)\Delta & l\Delta < x \leq (l+1)\Delta, l = -\frac{L}{2}+1\ldots\frac{L}{2}-2 \\ \frac{L-1}{2}\Delta & x \geq \left(\frac{L}{2}-1\right)\Delta \end{cases} \quad (4)$$

We use the Bussgang decomposition [9] to represent this non-linear function as a linear term plus an uncorrelated distortion term:

$$g(x) = \alpha x + \delta \quad (5)$$

where $\alpha$ is given by [9]:

$$\alpha = \frac{1}{\sigma_x^2} \int_{-\infty}^{\infty} x g(x) p(x) dx \quad (6)$$

where $p(x)$ is the probability density function (PDF) of the input signal $x$ (assumed zero mean), and $\sigma_x^2$ is its variance.

It is useful also to define, as in [11], a coefficient $\gamma$ relating the power of the quantized signal to that of the input signal, such that:

$$E\left[g^2(x)\right] = \gamma \sigma_x^2 = \alpha^2 \sigma_x^2 + \sigma_\delta^2 \quad (7)$$

We can then state a pair of expressions in closed form for these two parameters for the uniform quantizer, in the case of a Gaussian input:

*Proposition 1*: For the uniform quantizer defined in (4), with a zero mean Gaussian input variable of variance $\sigma_x^2$:

a) the linear scale factor $\alpha$ is given by:

$$\alpha = \frac{1}{\sqrt{2\pi}} \frac{\Delta}{\sigma_x} \left( \sum_{l=1}^{\frac{L}{2}-1} 2\exp\left(-\frac{l^2\Delta^2}{2\sigma_x^2}\right) + 1 \right) \quad (8), \text{ and}$$

b) the power scaling factor:

$$\gamma = \frac{\Delta^2}{\sigma_x^2}\left(\frac{1}{4} + 4\sum_{l=1}^{\frac{L}{2}-1} l Q\left(\frac{l\Delta}{\sigma_x}\right)\right) \quad (9)$$

where $Q(\cdot)$ denotes the Gaussian Q function.

*Proof*: See Appendix A

### B. Optimum Quantization

We next explore the optimum step size for the quantizer for given $L$ and hence fronthaul load. From (7) the quantization distortion power is given by:

$$\sigma_\delta^2 = \sigma_x^2\left(\gamma - \alpha^2\right) \quad (10)$$

Note that this implies that $\gamma \geq \alpha^2$. We observe, however, that it is not helpful to choose step size $\Delta$ in order to minimize this distortion power, since it is clear from (8)-(10) that $\sigma_\delta^2$ is minimized (to zero) for $\Delta = 0$, and that this also sets $\alpha$ to zero. Hence we instead minimize the signal to distortion noise ratio (SDNR) [11]:

$$\frac{\alpha^2 \sigma_x^2}{\sigma_\delta^2} = \frac{\alpha^2 \sigma_x^2}{\sigma_x^2(\gamma - \alpha^2)} = \frac{\alpha^2}{\gamma - \alpha^2} = \frac{1}{1 - \alpha^2/\gamma} \quad (11)$$

We can then find the optimum step size by solving the maximization problem:

$$P1: \Delta_{opt} = \arg\max_\Delta \left(\frac{1}{1-\alpha^2/\gamma}\right) = \arg\max_\Delta \left(\frac{\alpha^2}{\gamma}\right)$$

$$= \arg\max_\Delta \frac{\left(\frac{1}{2\pi}\frac{\Delta^2}{\sigma_x^2}\left(\sum_{l=1}^{\frac{L}{2}-1} 2\exp\left(-\frac{l^2\Delta^2}{2\sigma_x^2}\right)+1\right)^2\right)}{\frac{\Delta^2}{\sigma_x^2}\left(\frac{1}{4}+4\sum_{l=1}^{\frac{L}{2}-1} l Q\left(\frac{l\Delta}{\sigma_x}\right)\right)}$$

$$= \sigma_x \arg\max_{\Delta'} \frac{\left(\sum_{l=1}^{\frac{L}{2}-1} 2\exp\left(-\frac{l^2\Delta'^2}{2}\right)+1\right)^2}{\frac{1}{4}+4\sum_{l=1}^{\frac{L}{2}-1} l Q(l\Delta')} \quad (12)$$

While P1 does not result in a closed form expression for $\Delta$ (as noted in [10]), it can readily be solved numerically for given $L$, thanks to the form obtained in (12). The numerical results are the same as those listed in [10] for specific values of $L$, but (12) gives a straightforward method to obtain them for arbitrary $L$.

### C. Quantization in Uplink D-MaMIMO

The received uplink signal at AP $m$ is given by:

$$x_m = \sum_{k=1}^{K} g_{mk} s_k + n_m = \sum_{k=1}^{K} h_{mk}\sqrt{\beta_{mk}} s_k + n_m \quad (13)$$

The foregoing analysis assumes that the signal to be quantized is normally distributed. While this is not true if the path loss and shadowing is unknown, if $\beta_{mk}, \forall k$ is known at AP $m$, then the conditional probability density of the signal

$p_m(x_m|\beta_{mk}, \forall k)$ is normal, since the $h_{mk}$ terms and the noise are all normal. The variance is:

$$\sigma_m^2 = E\left[|x_m|^2\right] = \sum_{k=1}^{K} E\left[|h_{mk}|^2\right]\beta_{mk}\sigma_s^2 + \sigma_n^2 \\ = \sum_{k=1}^{K} \beta_{mk}\sigma_s^2 + \sigma_n^2 \quad (14)$$

This should clearly be used, along with the optimization of P1, to select the optimum quantization step size at each AP.

## IV. UPLINK PERFORMANCE OF D-MAMIMO

### A. Channel Estimation

In MaMIMO channel estimation is important, so we first discuss how the CPU estimates $g_{mk}$ from the quantized signals. We assume that the $k^{th}$ UT transmits a length $\tau$ pilot sequence $\sqrt{\tau}\boldsymbol{\phi}_k$, where the sequences $\boldsymbol{\phi}_k$ are orthonormal, that is $\boldsymbol{\phi}_k^H \boldsymbol{\phi}_{k'} = 1$ $k = k'$, or 0 otherwise. We first correlate the quantized signal from the $m^{th}$ AP with the pilot sequence $\boldsymbol{\phi}_k$ for the $k^{th}$ user:

$$r_{mk} = \boldsymbol{\phi}_k^H \mathbf{y}_m = \boldsymbol{\phi}_k^H (\alpha \mathbf{x}_m + \boldsymbol{\delta}_m) \\ = \boldsymbol{\phi}_k^H \alpha \sum_{k'=1}^{K} h_{mk'}\sqrt{\tau\beta_{mk'}}\boldsymbol{\phi}_{k'} + \alpha \boldsymbol{\phi}_k^H \mathbf{n}_m + \boldsymbol{\phi}_k^H \boldsymbol{\delta}_m \quad (15) \\ = \alpha h_{mk}\sqrt{\tau\beta_{mk}} + \alpha \boldsymbol{\phi}_k^H \mathbf{n}_m + \boldsymbol{\phi}_k^H \boldsymbol{\delta}_m$$

Hence we find a linear minimum mean square error (LMMSE) estimate of $g_{mk}$ by multiplying $r_{mk}$ by an optimized coefficient $c_{mk}$:

$$\hat{g}_{mk} = c_{mk} r_{mk} \quad (16)$$

*Proposition 2*: The optimum LMMSE estimator coefficient for $g_{mk}$ given $\beta_{mk}$ is:

$$c_{mk} = \frac{\beta_{mk}\sqrt{\tau}\alpha}{\tau\alpha^2 \beta_{mk} + (\gamma - \alpha^2)\sum_{k'=1}^{K}\beta_{mk'} + \gamma\sigma_n^2} \quad (17)$$

The mean square error (MSE) is then:

$$\sigma_e^2 = \frac{\beta_{mk}\left((\gamma - \alpha^2)\sum_{k'=1}^{K}\beta_{mk'} + \gamma\sigma_n^2\right)}{\alpha^2 \tau \beta_{mk} + (\gamma - \alpha^2)\sum_{k'=1}^{K}\beta_{mk'} + \gamma\sigma_n^2} \quad (18)$$

*Proof*: See Appendix B.

Comparing with (11) we note, however, that the MSE is minimized by choosing the quantization step $\Delta$ to maximize the SDNR, as discussed above.

### B. Data Detection

The quantized signal from AP $m$ can be written:

$$y_m = \alpha x_m + \delta_m = \alpha\left(\sum_{k=1}^{K} h_{mk}\sqrt{\beta_{mk}}s_k + n_m\right) + \delta_m \\ = \sum_{k=1}^{K} \alpha h_{mk}\sqrt{\beta_{mk}}s_k + \alpha n_m + \delta_m \quad (19)$$

We assume that the same number $b$ of bits per sample is used for the fronthaul connections to all APs, and hence the optimum normalized step size and hence $\alpha$ are the same for all. Then:

$$\mathbf{y} = \alpha\mathbf{Gs} + \alpha\mathbf{n} + \boldsymbol{\delta} \quad (20)$$

where $\boldsymbol{\delta}$ denotes the length $M$ vector of quantization distortions at the APs.

*Proposition 3*: We can then form an MMSE estimate of the transmitted data as:

$$\hat{\mathbf{s}} = \alpha\mathbf{G}^H \left(\alpha^2 \mathbf{GG}^H + \sigma_n^2 \mathbf{I} + \mathbf{C}_\delta\right)^{-1} \mathbf{y} \quad (21)$$

where the covariance matrix of $\boldsymbol{\delta}$:

$$\mathbf{C}_\delta = \text{diag}\left((\gamma - \alpha^2)\left(\sigma_s^2 \sum_{k=1}^{K}\beta_{mk} + \sigma_n^2\right), m=1\ldots M\right) \quad (22)$$

*Proof*: See Appendix C

## V. NUMERICAL RESULTS

In this section we give some numerical results for the performance of D-MaMIMO with quantized fronthaul, both in terms of channel estimation and of data detection, using linear MMSE estimation in both cases.

The channel elements, and in particular the values of $\beta_{mk}$ are obtained by Monte Carlo simulation, using the propagation model outlined above. The simulation parameters are as follows:

No. of users, $K = 40$

No. of APs, $M = 200$

Service area width, $l_{serv} = 1$ km

Edge signal to noise ratio, $SNR_{edge} = 20$ dB

The parameters of the path loss model are given in section II above.

We then obtain the MSE of the channel estimate using (18): however it is more useful to report a normalized MSE as a fraction of the corresponding mean square channel gain, $\beta_{mk}$, and the cumulative probability distribution function of this is given in Fig. 2.

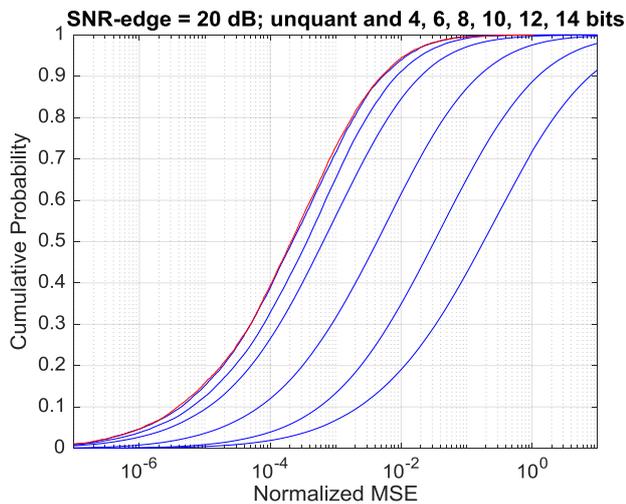

Fig. 2. CDF of normalized MSE of channel estimate with 4,6,8,10,12 and 14 bits quantization of fronthaul (blue lines, right to left), compared to unquantized fronthaul (red)

estimation and data detection at the central processing unit, making use of closed form expressions arising from the Bussgang decomposition. The expressions derived have also been verified by direct simulations. Numerical results are given indicating the quantization can have a very significant effect on both channel estimation and data detection, and that very high levels of quantization may be required to obtain the performance available in an unquantized system, as indicated by the distribution of the SINR.

The results given for data detection do not include the effects of channel estimation error: further work is required to evaluate this effect. There is also further work possible on more advanced optimization approaches, especially robust optimization taking into account the unreliability of channel estimates. The complexity and scalability of the detection approaches should also be addressed.

We note that a very large resolution is required to achieve the accuracy provided by the unquantized backhaul. For this value of edge SNR the unquantized performance is only approached with 14 bit quantization.

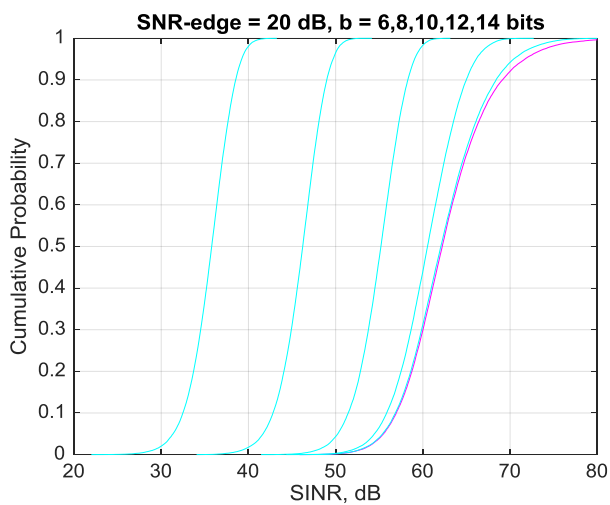

Fig. 3. Signal to interference plus noise and quantization distortion ratio with unquantized (magenta) and quantized (cyan) fronthaul with 6-14 quantization bits (left to right)

A similar result is shown in Fig. 3 for the overall SINR (including quantization distortion). We note that in this case perfect channel estimation is assumed. Again it is clear that a large fronthaul capacity is required to achieve the full capacity available to the network.

Note that the results obtained here from the expressions derived above have also been verified by comparison with direct simulations.

## VI. CONCLUSIONS

We have analyzed the performance of a distributed massive MIMO system with quantized backhaul, using MMSE channel

APPENDIX A

Use (4) and (6) to find $\alpha$ for uniform quantizer:

$$\alpha = \frac{1}{\sigma_x^2} \int_{-\infty}^{\infty} x g(x) p_N(x,\sigma_x) dx$$

$$= \frac{1}{\sigma^2} \begin{pmatrix} \int_{-\infty}^{-\frac{L}{2}+1} -x \frac{L-1}{2} \Delta\, p_N(x,\sigma_x) dx \\ + \sum_{l=-\frac{L}{2}+1}^{\frac{L}{2}-2} \int_{l\Delta}^{(l+1)\Delta} x\left(l+\frac{1}{2}\right)\Delta\, p_N(x,\sigma_x) dx \\ + \int_{\frac{L}{2}-1}^{\infty} x \frac{L-1}{2}\Delta\, p_N(x,\sigma_x) dx \end{pmatrix}$$

$$= \frac{(L-1)}{2\sqrt{2\pi}} \frac{\Delta}{\sigma_x} \exp\left(-\frac{\left(\frac{L}{2}-1\right)^2}{2\sigma_x^2}\right)$$

$$+ \sum_{l=-\frac{L}{2}+1}^{\frac{L}{2}-2} \frac{\left(l+\frac{1}{2}\right)}{\sqrt{2\pi}} \frac{\Delta}{\sigma_x} \begin{pmatrix} \exp\left(-\frac{l^2\Delta^2}{2\sigma_x^2}\right) \\ -\exp\left(-\frac{(l+1)^2\Delta^2}{2\sigma_x^2}\right) \end{pmatrix}$$

$$+ \frac{(L-1)}{2\sqrt{2\pi}} \frac{\Delta}{\sigma_x} \exp\left(-\frac{\left(\frac{L}{2}-1\right)^2}{2\sigma_x^2}\right)$$

$$= \sum_{l=-\frac{L}{2}+1}^{\frac{L}{2}-1} \frac{\left(l+\frac{1}{2}\right)}{\sqrt{2\pi}} \frac{\Delta}{\sigma_x} \exp\left(-\frac{l^2\Delta^2}{2\sigma_x^2}\right)$$

$$- \sum_{l'=-\frac{L}{2}+1}^{\frac{L}{2}-1} \frac{\left(l'-\frac{1}{2}\right)}{\sqrt{2\pi}} \frac{\Delta}{\sigma_x} \exp\left(-\frac{l'^2\Delta^2}{2\sigma_x^2}\right)$$

$$= \sum_{l=-\frac{L}{2}+1}^{\frac{L}{2}-1} \frac{1}{\sqrt{2\pi}} \frac{\Delta}{\sigma_x} \exp\left(-\frac{l^2\Delta^2}{2\sigma_x^2}\right) \quad (23)$$

$$= \sum_{l=1}^{\frac{L}{2}-1} \frac{2}{\sqrt{2\pi}} \frac{\Delta}{\sigma_x} \exp\left(-\frac{l^2\Delta^2}{2\sigma_x^2}\right) + \frac{1}{\sqrt{2\pi}} \frac{\Delta}{\sigma_x}$$

$$= \frac{1}{\sqrt{2\pi}} \frac{\Delta}{\sigma_x} \left( \sum_{l=1}^{\frac{L}{2}-1} 2\exp\left(-\frac{l^2\Delta^2}{2\sigma_x^2}\right) + 1 \right)$$

Here $l' = l+1$

Next we find $\gamma$ using:

$$\gamma = \frac{E\left[g^2(x)\right]}{E\left[x^2\right]} = \frac{1}{\sigma_x^2} \int_{-\infty}^{\infty} g^2(x) p_x(x) dx$$

$$= \frac{2}{\sigma_x^2} \int_0^{\infty} g^2(x) p_N(x,\sigma_x) dx$$

$$= \frac{2}{\sigma_x^2} \begin{pmatrix} \sum_{l=1}^{\frac{L}{2}-1} \int_{(l-1)\Delta}^{l\Delta} \left(l-\frac{1}{2}\right)^2 \Delta^2\, p_N(x,\sigma_x) dx \\ + \int_{\left(\frac{L}{2}-1\right)\Delta}^{\infty} \left(\frac{L-1}{2}\right)^2 \Delta^2\, p_N(x,\sigma_x) dx \end{pmatrix}$$

$$= \frac{2\Delta^2}{\sigma_x^2} \begin{pmatrix} \sum_{l=1}^{\frac{L}{2}-1} \left(l-\frac{1}{2}\right)^2 \left(Q\left(\frac{(l-1)\Delta}{\sigma_x}\right) - Q\left(\frac{l\Delta}{\sigma_x}\right)\right) \\ + \left(\frac{L-1}{2}\right)^2 Q\left(\frac{(L-2)\Delta}{2\sigma_x}\right) \end{pmatrix}$$

$$= \frac{2\Delta^2}{\sigma_x^2} \begin{pmatrix} \sum_{l=1}^{\frac{L}{2}-1} \left(l-\frac{1}{2}\right)^2 Q\left(\frac{(l-1)\Delta}{\sigma_x}\right) - \sum_{l=1}^{\frac{L}{2}-1} \left(l-\frac{1}{2}\right)^2 Q\left(\frac{l\Delta}{\sigma_x}\right) \\ + \left(\frac{L-1}{2}\right)^2 Q\left(\frac{(L-2)\Delta}{2\sigma_x}\right) \end{pmatrix}$$

$$= \frac{2\Delta^2}{\sigma_x^2} \left[ \sum_{l'=0}^{\frac{L}{2}-1} \left(l'+\frac{1}{2}\right)^2 Q\left(\frac{l'\Delta}{\sigma_x}\right) - \sum_{l=1}^{\frac{L}{2}-1} \left(l-\frac{1}{2}\right)^2 Q\left(\frac{l\Delta}{\sigma_x}\right) \right]$$

$$= \frac{2\Delta^2}{\sigma_x^2} \left[ \left(\frac{1}{2}\right)^2 Q(0) + \sum_{l=1}^{\frac{L}{2}-1} \left(\left(l+\frac{1}{2}\right)^2 - \left(l-\frac{1}{2}\right)^2\right) Q\left(\frac{l\Delta}{\sigma_x}\right) \right] \quad (24)$$

$$= \frac{\Delta^2}{\sigma_x^2} \left[ \frac{1}{4} + 4\sum_{l=1}^{\frac{L}{2}-1} l\, Q\left(\frac{l\Delta}{\sigma_x}\right) \right]$$

In this case $l' = l-1$.

APPENDIX B

The mean square error of the channel estimate (using (10), (14), (15) and (16) and assuming unit signal power $\sigma_s^2 = 1$):

$$\sigma_e^2 = \mathrm{E}\left[|\hat{g}_{mk} - g_{mk}|^2\right] = \mathrm{E}\left[|c_{mk} r_{mk} - h_{mk}\sqrt{\beta_{mk}}|^2\right]$$

$$= \mathrm{E}\left[\left|h_{mk}\sqrt{\beta_{mk}}\left(c_{mk}\sqrt{\tau}\alpha - 1\right) + c_{mk}\alpha\,\boldsymbol{\phi}_k^H\mathbf{n} + c_{mk}\boldsymbol{\phi}_k^H\boldsymbol{\delta}_m\right|^2\right]$$

$$\stackrel{(a)}{=} \beta_{mk}\left(c_{mk}\sqrt{\tau}\alpha - 1\right)^2 \mathrm{E}\left[|h_{mk}|^2\right] + c_{mk}^2\alpha^2\sigma_n^2 + c_{mk}^2\sigma_{\delta,m}^2$$

$$= \beta_{mk}\left(c_{mk}\sqrt{\tau}\alpha - 1\right)^2 + c_{mk}^2\alpha^2\sigma_n^2 + c_{mk}^2\left(\gamma - \alpha^2\right)\sigma_m^2 \quad (25)$$

$$= \beta_{mk}\left(c_{mk}\sqrt{\tau}\alpha - 1\right)^2$$
$$+ c_{mk}^2\left(\left(\gamma - \alpha^2\right)\left(\sum_{k'=1}^{K}\beta_{mk'}\sigma_s^2 + \sigma_n^2\right) + \alpha^2\sigma_n^2\right)$$

$$= \beta_{mk}\left(c_{mk}\sqrt{\tau}\alpha - 1\right)^2 + c_{mk}^2\left(\left(\gamma - \alpha^2\right)\sum_{k'=1}^{K}\beta_{mk'} + \gamma\sigma_n^2\right)$$

Note that equality (*a*) depends on the assumption, using the Bussgang decomposition for a non-linear function $g(x+z) = \alpha(x+z) + d$, where $x$ and $z$ are uncorrelated zero mean Gaussian variables, and the Bussgang coefficient $\alpha$ is calculated assuming input variance $\sigma_x^2 + \sigma_z^2$, that the distortion term $d$ is uncorrelated with both $x$ and $z$. By definition we have that $x + z$ is uncorrelated with $d$: but it does not follow that the variables are individually uncorrelated. However, given that $d$ is in fact a deterministic function $d(x+z)$ of its argument it is possible to show numerically that $\iint_{x,z} p_x(x) p_z(z) x\, d(x+z)\, dx\, dz$ is zero to high accuracy, from which it follows that $z$ is also uncorrelated with $d$.

Find optimum estimator coefficient $c_{mk}$ using:

$$\frac{\partial \sigma_e^2}{\partial c_{mk}} = 0 \rightarrow$$

$$2\beta_{mk}\sqrt{\tau}\alpha\left(c_{mk}\sqrt{\tau}\alpha - 1\right)$$
$$+ 2c_{mk}\left(\left(\gamma - \alpha^2\right)\sum_{k'=1}^{K}\beta_{mk'} + \gamma\sigma_n^2\right) = 0 \quad (26)$$

$$c_{mk}\beta_{mk}\tau\alpha^2 + c_{mk}\left(\left(\gamma - \alpha^2\right)\sum_{k'=1}^{K}\beta_{mk'} + \gamma\sigma_n^2\right) = \beta_{mk}\sqrt{\tau}\alpha$$

$$c_{mk,opt} = \frac{\beta_{mk}\sqrt{\tau}\alpha}{\tau\alpha^2\beta_{mk} + \left(\gamma - \alpha^2\right)\sum_{k'=1}^{K}\beta_{mk'} + \gamma\sigma_n^2}$$

Then substitute in (25) to give:

$$\sigma_e^2 = \beta_{mk}\left(\frac{\beta_{mk}\sqrt{\tau}\alpha}{\alpha^2\tau\beta_{mk} + (\gamma - \alpha^2)\sum_{k'=1}^{K}\beta_{mk'} + \gamma\sigma_n^2}\sqrt{\tau}\alpha - 1\right)^2$$

$$+ \left(\frac{\beta_{mk}\sqrt{\tau}\alpha}{\alpha^2\tau\beta_{mk} + (\gamma-\alpha^2)\sum_{k'=1}^{K}\beta_{mk'}+\gamma\sigma_n^2}\right)^2 \left((\gamma-\alpha^2)\sum_{k'=1}^{K}\beta_{mk'}+\gamma\sigma_n^2\right)$$

$$= \frac{\beta_{mk}\left((\gamma-\alpha^2)\sum_{k'=1}^{K}\beta_{mk'}+\gamma\sigma_n^2\right)^2}{\left(\alpha^2\tau\beta_{mk}+(\gamma-\alpha^2)\sum_{k'=1}^{K}\beta_{mk'}+\gamma\sigma_n^2\right)^2}$$

$$+ \frac{\left(\beta_{mk}\sqrt{\tau}\alpha\right)^2\left((\gamma-\alpha^2)\sum_{k'=1}^{K}\beta_{mk'}+\gamma\sigma_n^2\right)}{\left(\alpha^2\tau\beta_{mk}+(\gamma-\alpha^2)\sum_{k'=1}^{K}\beta_{mk'}+\gamma\sigma_n^2\right)^2}$$

$$= \frac{\beta_{mk}\left((\gamma-\alpha^2)\sum_{k'=1}^{K}\beta_{mk'}+\gamma\sigma_n^2\right)\left(\left((\gamma-\alpha^2)\sum_{k'=1}^{K}\beta_{mk'}+\gamma\sigma_n^2\right)+\alpha^2\tau\beta_{mk}\right)}{\left(\alpha^2\tau\beta_{mk}+(\gamma-\alpha^2)\sum_{k'=1}^{K}\beta_{mk'}+\gamma\sigma_n^2\right)^2} \quad (27)$$

$$= \frac{\beta_{mk}\left((\gamma-\alpha^2)\sum_{k'=1}^{K}\beta_{mk'}+\gamma\sigma_n^2\right)}{\alpha^2\tau\beta_{mk}+(\gamma-\alpha^2)\sum_{k'=1}^{K}\beta_{mk'}+\gamma\sigma_n^2}$$

APPENDIX C

From (20) we have:

$$\mathbf{y} = \alpha\mathbf{Gs} + \alpha\mathbf{n} + \boldsymbol{\delta} \quad (28)$$

and hence (using the orthogonality principle of MMSE estimation) that the optimum weight matrix:

$$\mathbf{W} = \mathrm{E}\left[\mathbf{s}\mathbf{y}^H\right]\mathrm{E}\left[\mathbf{y}\mathbf{y}^H\right]^{-1}$$

$$\mathrm{E}\left[\mathbf{s}\mathbf{y}^H\right] = \alpha\mathrm{E}\left[\mathbf{s}\mathbf{s}^H\right]\mathbf{G}^H + \alpha\mathrm{E}\left[\mathbf{s}\mathbf{n}^H\right] + \mathrm{E}\left[\mathbf{s}\boldsymbol{\delta}\right] = \alpha\mathbf{G}^H$$

$$\mathrm{E}\left[\mathbf{y}\mathbf{y}^H\right] = (\alpha\mathbf{Gs}+\alpha\mathbf{n}+\boldsymbol{\delta})(\alpha\mathbf{Gs}+\alpha\mathbf{n}+\boldsymbol{\delta})^H \quad (29)$$

$$= \alpha^2\mathrm{E}\left[\mathbf{s}\mathbf{s}^H\right]\mathbf{G}\mathbf{G}^H + \alpha^2\mathrm{E}\left[\mathbf{n}\mathbf{n}^H\right] + \mathbf{C}_{\boldsymbol{\delta}}$$

$$= \alpha^2\mathbf{G}\mathbf{G}^H + \alpha^2\sigma_n^2 + \mathbf{C}_{\boldsymbol{\delta}}$$

Here we use the same result as in Appendix B above that $\boldsymbol{\delta}$ is uncorrelated with both $\mathbf{s}$ and $\mathbf{n}$, and also that $\mathbf{s}$ and $\mathbf{y}$ are both zero mean. Then:

$$\mathbf{C}_{\boldsymbol{\delta}} = \mathrm{E}\left[\boldsymbol{\delta}\boldsymbol{\delta}^H\right] \stackrel{(a)}{=} \mathrm{diag}\left(\mathrm{E}\left[|\delta_m|^2\right], m=1\ldots M\right)$$
$$\stackrel{(b)}{=} \mathrm{diag}\left((\gamma-\alpha^2)\left(\sum_{k=1}^{K}\beta_{mk}+\sigma_n^2\right), m=1\ldots M\right) \quad (30)$$

where (*a*) arises because the elements of the distortion vector are uncorrelated, and (*b*) follows from (10). Then:

$$\hat{\mathbf{s}} = \mathbf{W}\mathbf{y} = \mathbf{G}^H\left(\alpha^2\mathbf{G}\mathbf{G}^H + \alpha^2\sigma_n^2 + \mathbf{C}_{\boldsymbol{\delta}}\right)\mathbf{y} \quad (31)$$

and the error:

$$\begin{aligned}
\mathbf{e} &= \hat{\mathbf{s}} - \mathbf{s} = \mathbf{W}\mathbf{y} - \mathbf{s} \\
&= (\alpha\mathbf{W}\mathbf{G} - \mathbf{I})\mathbf{s} + \mathbf{W}(\alpha\mathbf{n} + \boldsymbol{\delta})
\end{aligned}$$

The covariance matrix of the error (for given $\mathbf{G}$) is then given by:

$$\begin{aligned}
\mathbf{C}_e &= \mathop{\mathrm{E}}_{\mathbf{s},\mathbf{n}}\left[\mathbf{e}\mathbf{e}^H\right] = \mathop{\mathrm{E}}_{\mathbf{s},\mathbf{n}}\left[\left((\alpha\mathbf{W}\mathbf{G} - \mathbf{I})\mathbf{s} + \mathbf{W}(\alpha\mathbf{n} + \boldsymbol{\delta})\right)\left((\alpha\mathbf{W}\mathbf{G} - \mathbf{I})\mathbf{s} + \mathbf{W}(\alpha\mathbf{n} + \boldsymbol{\delta})\right)^H\right] \\
&= (\alpha\mathbf{W}\mathbf{G} - \mathbf{I})(\alpha\mathbf{W}\mathbf{G} - \mathbf{I})^H + \alpha^2\sigma_n^2 \mathbf{W}\mathbf{W}^H + \mathbf{W}\mathbf{C}_{\boldsymbol{\delta}}\mathbf{W}^H \\
&= \alpha^2 \mathbf{W}\mathbf{G}\mathbf{G}^H\mathbf{W}^H - \alpha\left(\mathbf{W}\mathbf{G} + (\mathbf{W}\mathbf{G})^H\right) + \mathbf{I} + \alpha^2\sigma_n^2 \mathbf{W}\mathbf{W}^H + \mathbf{W}\mathbf{C}_{\boldsymbol{\delta}}\mathbf{W}^H \\
&= \mathbf{W}\left(\alpha^2\mathbf{G}\mathbf{G}^H + \alpha^2\sigma_n^2 + \mathbf{C}_{\boldsymbol{\delta}}\right)\mathbf{W}^H - 2\alpha\mathbf{W}\mathbf{G} + \mathbf{I} \\
&= \sigma_s^2\mathbf{I} - \alpha\mathbf{G}^H\left(\alpha^2\mathbf{G}\mathbf{G}^H\sigma_s^2 + \mathbf{C}_{\boldsymbol{\delta}} + \alpha^2\sigma_n^2\mathbf{I}\right)^{-1}\alpha\mathbf{G} \\
&= \sigma_s^2\mathbf{I} - \left(\sigma_s^2\mathbf{I} + \frac{1}{\alpha^2}\left(\mathbf{G}^H\mathbf{C}_{\boldsymbol{\delta}}^{-1}\mathbf{G}\right)^{-1} + \sigma_n^2\left(\mathbf{G}^H\mathbf{G}\right)^{-1}\right)^{-1}
\end{aligned} \quad (32)$$

We next determine the expected value of the error over random $h_{mk}$. Considering the third term in the bracket of (32), the off-diagonal elements of $\mathrm{E}\left[\left(\mathbf{G}^H\mathbf{G}\right)^{-1}\right]$ are clearly zero for uncorrelated Rayleigh fading, while the diagonal elements are given by:

$$\mathrm{E}\left[\left(\mathbf{G}^H\mathbf{G}\right)^{-1}\right]_{kk} \geq \frac{1}{\mathrm{E}\left[\mathbf{G}^H\mathbf{G}\right]_{kk}} = \frac{1}{\sum_{m=1}^M \mathrm{E}\left[g_{mk}g_{mk}^*\right]} = \frac{1}{\sum_{m=1}^M \beta_{mk}} \quad (33)$$

$$\mathrm{E}\left[\left(\mathbf{G}^H\mathbf{C}_{\boldsymbol{\delta}}^{-1}\mathbf{G}\right)^{-1}\right]_{kk'} \geq \frac{1}{\sum_{m=1}^M \mathrm{E}\left[g_{mk}c_{\delta,m}^{-1}g_{mk'}^*\right]}
= \begin{cases} \dfrac{1}{\sum_{m=1}^M \beta_{mk}c_{\delta,m}^{-1}} & k = k' \\ 0 & k \neq k' \end{cases} \quad (34)$$

where:

$$c_{\delta,m} = (\gamma - \alpha^2)\left(\sum_{k=1}^K \beta_{mk} + \sigma_n^2\right) \quad (35)$$

denotes the $m^{\mathrm{th}}$ entry of the diagonal of $\mathbf{C}_{\boldsymbol{\delta}}$, and:

$$\mathrm{E}\left[\left(\mathbf{G}^H\mathbf{C}_{\boldsymbol{\delta}}^{-1}\mathbf{G}\right)^{-1}\right]_{kk} \geq \frac{\gamma - \alpha^2}{\sum_{m=1}^M \beta_{mk} \Big/ \left(\sum_{k'=1}^K \beta_{mk'} + \sigma_n^2\right)} \quad (36)$$